# Energy-based Graph Convolutional Networks for Scoring Protein Docking Models

(Running Title: EGCN for Scoring in Protein Docking)

Yue Cao[1] and Yang Shen[1,2]*

[1] Department of Electrical and Computer Engineering, [2] TEES-AgriLife Center for Bioinformatics and Genomic Systems Engineering, Texas A&M University, College Station, Texas 77843, USA

* Correspondence: yshen@tamu.edu

**Abstract:**
Structural information about protein-protein interactions, often missing at the interactome scale, is important for mechanistic understanding of cells and rational discovery of therapeutics. Protein docking provides a computational alternative to predict such information. However, ranking near-native docked models high among a large number of candidates, often known as the scoring problem, remains a critical challenge. Moreover, estimating model quality, also known as the quality assessment problem, is rarely addressed in protein docking.

In this study the two challenging problems in protein docking are regarded as relative and absolute scoring, respectively, and addressed in one physics-inspired deep learning framework. We represent proteins and encounter complexes as intra- and inter-molecular residue contact graphs with atom-resolution node and edge features. And we propose a novel graph convolutional kernel that pool interacting nodes' features through edge features so that generalized interaction energies can be learned directly from graph data. The resulting energy-based graph convolutional networks (EGCN) with multi-head attention are trained to predict intra- and inter-molecular energies, binding affinities, and quality measures (interface RMSD) for encounter complexes. Compared to a state-of-the-art scoring function for model ranking, EGCN has significantly improved ranking for a CAPRI test set involving homology docking; and is comparable for Score_set, a CAPRI benchmark set generated by diverse community-wide docking protocols not known to training data. For Score_set quality assessment, EGCN shows about 27% improvement to our previous efforts. Directly learning from structure data in graph representation, EGCN represents the first successful development of graph convolutional networks for protein docking.

# INTRODUCTION

Protein-protein interactions (PPIs) underlie many important cellular processes. Structural information about these interactions often helps reveal their mechanisms, understand diseases, and develop therapeutics. However, such information is often unavailable at the scale of protein-protein interactomes [1], which calls for computational protein docking methods. Two major tasks for protein docking are sampling (aiming at generating many near-native models or decoys) and scoring (aiming at identifying those near-natives among a large number of candidate decoys), both of which present tremendous challenges [2].

This study focuses on scoring by filling two gaps in protein docking. First, current scoring functions for protein docking often aim at relative scoring, in other words, ranking near-natives high [2]. But they do not score the decoy/model quality directly (absolute scoring), which is more often known as quality estimation or quality assessment (without known native structures) in the protein-structure community of CASP [3]. Although quality estimation methods based on machine learning are emerging for single-protein structure prediction[4–9], such methods are still rare for protein-complex structure prediction, i.e., protein docking [10]. Second, state-of-the-art scoring functions (for relative scoring) in protein docking are often based on machine learning with hand-engineered features, such as physical-energy terms [11, 12], statistical potentials [13, 14], and graph kernels [15]. These features, heavily relying on domain expertise, are often not specifically tailored or optimized for scoring purposes. Recently, deep learning has achieved tremendous success in image recognition and natural language processing [16, 17], largely due to its automated feature/representation learning from raw inputs of image pixels or words. This trend has also rippled to structural bioinformatics and will be briefly reviewed later.

To fill the aforementioned gaps for scoring in protein docking, we propose a deep learning framework with automated feature learning to estimate the quality of protein-docking models (measured by interface RMSD or iRMSD) and to rank them accordingly. In other words, the framework simultaneously addresses both relative scoring (ranking) and absolute scoring (quality assessment) using features directly learned from data. To that end, our deep learning framework predicts binding free energy values of docking models (encounter complexes) whereas these values are correlated to known binding free energy values of native complexes according to model quality.

Technical challenges remain for the framework of deep learning: how to represent protein-complex structure data and learn from such data effectively? Current deep learning methods in structural bioinformatics often use 3D volumetric representations of molecular structures. For instance, for protein-structure quality assessment, atom density maps have been used as the raw input to 3D Convolutional Neural Networks (CNN) [18, 19]. For protein-ligand interaction prediction, the Atomic Convolutional Neural Network (ACNN) [20] uses the neighbor matrix as input and uses radical pooling to simulate the additive pairwise interaction. And for RNA structure QA, grid representations of the structures are used as the input to 3D CNN [21]. However, learning features from volumetric data of molecular structures present several drawbacks [22]. First, representing the volumetric input data as pixel data through tensors would require discretization, which may lose some biologically-meaningful features while costing time. Second, such input data are often sparse, resulting in many convolutional operations for zero-valued pixels and thus low efficiency. Third, the convolutional operation is not rotation-invariant, which demands rotational augmentation of training data and increased computational cost.

To effectively learn features from and predict labels for protein-docking structure models, we represent proteins and protein-complexes as intra- and inter-molecular residue contact graphs with atom-resolution node and edge features. Such a representation naturally captures the spatial relationship of protein-complex structures while overcoming the aforementioned drawbacks of learning from volumetric data. Moreover, we learn from such graph data by proposing a physics-inspired graph convolutional kernel that pool interacting nodes' interactions through their node and edge features. We use two resulting energy-based graph convolutional networks (EGCN) of the same architecture but different parameters to predict intra- and inter-molecular energy potentials for protein-docking models (encounter complexes), which further predicts these encounter complexes' binding affinity and quality (iRMSD) values. Therefore, our EGCN is capable of both model ranking and quality estimation (or quality assessment without known native structures). We note the recent surging of graph neural networks in protein modeling, such as protein interface prediction [23] and protein structure classification [22]. We also note that this is the first study of graph neural networks for protein docking.

The rest of the paper is organized as follows. In the Methods section, we first describe the data sets for training, validating, and testing our EGCN models. We proceed to introduce contact graph representations for proteins and protein complexes and initial features for nodes and edges; our energy-based graph convolutional networks as the intra- or inter-molecular energy predictor for encounter complexes; and additional contents (label, loss function, and optimization) for training our EGCN. Next we compare in Results EGCN's performances against a state-of-the-art scoring function for ranking and against our previous efforts for both ranking and quality estimation. We also share in Discussion our thoughts on how EGCN can be further improved in accuracy and training efficiency. Lastly, we summarize major contributions in method development and major results in method performances in Conclusion.

## MATERIALS AND METHODS

### Data

*Training and Validation Sets*

We randomly choose 50 protein-protein pairs from protein benchmark 4.0[24] as in our earlier work [10] and split them into training and validation pairs with the ratio 4:1. For each training or validation pair, we perform rigid docking using ClusPro[25], retain the top-1000 decoys according to ClusPro's default scoring, and introduce flexible perturbation for each decoy using cNMA[10, 26, 27]. In total, we have 40,000 training and 10,000 validation decoys.

*Test Sets*

We consider three test sets of increasing difficulty levels.

1. The first test set includes the rest 107 protein pairs from the protein benchmark set 4.0[24]. As in the training and validation sets, these unbound protein pairs are rigidly docked through ClusPro. The top 1000 decoys after scoring are flexibly perturbed through cNMA, leading to 107,000 decoys.
2. This second test set includes 14 recent CAPRI targets[10] undergoing the same protocol above (ClusPro + cNMA) for 14,000 decoys. Besides unbound docking seen in the benchmark test set, many of the targets here involve homology-unbound and homology-homology docking[26].
3. The last test set is Score_set, a CAPRI benchmark for scoring[28]. It includes 13 earlier CAPRI targets, each of which has 400 to 2,000 flexible decoys generated through various protocols by the community.

We summarize PDB IDs for training, validation, and the three test sets in the supplementary Table S1.

**Graphs and Features**

The structure of a protein or protein complex is represented as a graph $\mathcal{G}$ where each residue corresponds to a node and each intra- or inter-molecular residue-pair defines an edge. Such intra- and inter-molecular contact graphs can be united under bipartite graphs: the two sets of nodes correspond to the same protein for intra-molecular graphs and they do to binding partners for inter-molecular graphs. Atom-resolution features are further introduced for nodes and edges of such bipartite contact graphs. In other words, to represent $\mathcal{G}$, we have initialized two node feature matrices (denoted by $X_A \in R^{N_1 \times M}$ and $X_B \in R^{N_2 \times M}$) and an edge feature tensor (denoted by $A \in R^{N_1 \times N_2 \times K}$), where $N_1$ or $N_2$ denote the number of residues for either protein $A$ or $B$ (which can be the same protein in the case of intra-molecular contact graphs), $M$ the number of features per node, and $K$ the number of features per edge. As revealed later in our graph convolutional networks, node feature matrices will be learned, i.e. updated layer after layer, by interacting with neighboring nodes' features through the fixed edge tensors.

For node-feature matrix $X$, we use $M = 4$ features for each node. Each side chain is represented as a pseudo atom whose position is the geometric center of the side chain. And the first 3 node features are the side-chain pseudo atom's charge, non-bonded radii, and distance-to-$C_\alpha$, as parameterized in the Rosetta coarse-grained energy model [29]. The last node feature is the solvent accessible surface area (SASA) of the whole residue, as calculated by the FreeSASA program [30].

For edge-feature tensor $A$, we use $K = 11$ features for each edge. These features are related to pairwise atomic distances between the two corresponding residues. Specifically, each residue is a point cloud of 6 atoms, including 5 backbone atoms (N, HN, CA, C, and O as named in CHARMM27) and 1 side-chain pseudo atom (named SC). For numerical efficiency, we have picked 11 pairwise distances (including potential hydrogen bonds) out of the total $6 \times 6 = 36$ (see Table I) and used their reciprocals for the edge features. We use a cutoff of 12 Å for pairwise atomic distances and set edge features at zero when corresponding distances are above the cutoff.

**Table I**. The atom pairs whose distance are converted to edge features. Atom names follow the convention in CHARMM, except that "SC" corresponds to pseudo-atoms for side chains. Note that the last two features are set at 0 for pairs involving prolines (without HN).

| Index | Atom 1 | Atom 2 |
|-------|--------|--------|
| 1 | SC | SC |
| 2 | SC | O |
| 3 | SC | N |
| 4 | O | SC |
| 5 | O | O |
| 6 | O | N |
| 7 | N | SC |
| 8 | N | O |
| 9 | N | N |
| 10 | HN | O |
| 11 | O | HN |

In this study, in order to address input proteins of varying sizes, we consider the maximum number of residues to be 500 and hence fix $N$ at 500 for contact graphs. Correspondingly, smaller proteins' feature matrices/tensors $X$ and $A$ are zero-padded as needed so that feature matrices/tensors for all input data are of the same sizes: $X \in R^{N \times M}$ and $A \in R^{N \times N \times K}$.

## Energy-based Graph Convolutional Networks

*Background on principle-driven energy model*

To score protein-docking models, we try to use machine learning to model $\Delta G$, the binding energy of encounter complexes, which can be written as:

$$\Delta G = G_C - G_{Ru} - G_{Lu}$$

where $G_C$, $G_{Ru}$, and $G_{Lu}$ denote the Gibbs free energies of the complex, unbound receptor and unbound ligand, respectively. For $G_C$, we can further decompose it into the intra-molecular energies within two individual proteins and the inter-molecular energy between two proteins:

$$G_C = G_R + G_L + G_{RL}$$

where $G_R$, $G_L$, $G_{RL}$ are the Gibbs free energies within the encountered receptor, within the encountered ligand and between them, respectively. Therefore, the binding energy in classical force fields can be written as:

$$\Delta G = G_R + G_L - G_{Ru} - G_{Lu} + G_{RL}$$

In flexible docking, unbound and encountered structures of the same protein are often different. Such protein conformational changes indicate that $G_R - G_{Ru} \neq 0$ and $G_L - G_{Lu} \neq 0$.

*Extension to data-driven energy model*

It is noteworthy that the expression of $\Delta G$ above consists of four terms measuring intra-molecular free energies of individual proteins and one one term measuring inter-molecular free energies across two proteins. Therefore, we decide to use two machine-learning models $f$ of the same neural-network architecture but different parameters to approximate the two types of energy terms as follows:

$$\Delta G \approx \Delta \hat{G} = f_\theta(X_R, X_R, A_{RR}) + f_\theta(X_L, X_L, A_{LL}) - f_\theta(X_{Ru}, X_{Ru}, A_{RuRu}) - f_\theta(X_{Lu}, X_{Lu}, A_{LuLu}) + f_{\theta'}(X_R, X_L, A_{RL})$$

where $f_\theta$ and $f_{\theta'}$ are the intra- and inter-molecular energy models (graph convolutional networks here) parameterized by $\theta$ and $\theta'$, respectively. Subscripts are included for node feature matrices $X$ or edge feature matrices $A$ to indicate identities of molecules or molecular pairs. The parameters are to be learned from data specifically for the purpose of quality estimation.

The architecture of the neural networks is summarized in Figure 1 whereas individual components are detailed below.

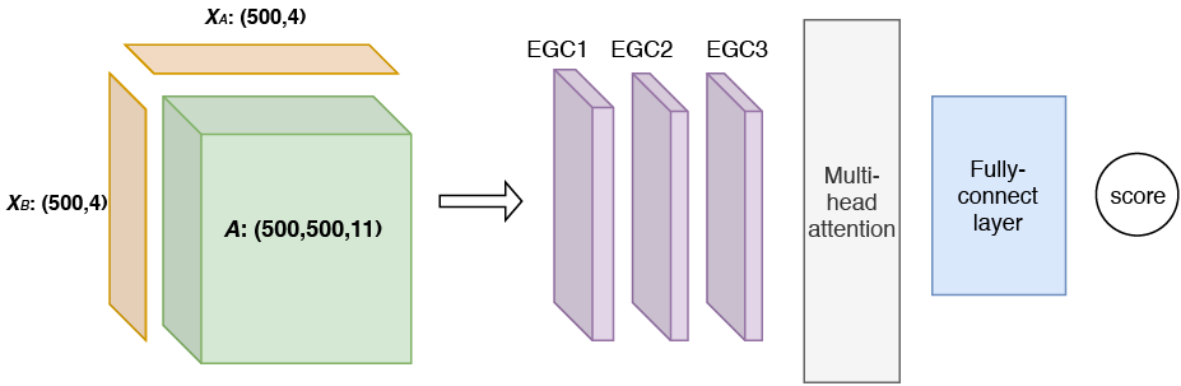

**Figure 1.** The architecture of the proposed graph convolutional network (GCN) models for intra- or inter-molecular energies. In our work, there are five types of such models together for predicting encounter-complex binding energy, including 4 intra-molecular models with shared parameters for the unbound or encountered receptor or ligand as well as 1 inter-molecular model for the encounter complex. In each type of model, the inputs (to the left of the arrow) include a pair of node-feature matrices ($X_A$ and $X_B$) for individual protein(s) and an edge-feature tensor ($A$) for intra- or inter-molecular contacts. And the inputs are fed through 3 layers of our energy-based graph convolution layers that learn from training data to aggregate and transform atomic interactions, followed by multi-head attention module and fully-connected layers for the output of intra- or inter-molecular energy.

*Energy-based Graph Convolutional layer (EGC)*
The node matrices ($X_A$ and $X_B$) and the edge tensor $A$ of each contact graph are first fed to three consecutive graph convolutional layers inspired by physics. Specifically, energy potentials are often pairwise additive and those between atoms ($i$ and $j$) are often of the form $\frac{x_i^T x_j}{r^a}$, where $x_i$ and $x_j$ are atom "features", $r$ is the distance between them, and $a$ is an integer. For instance, $x$ being a scalar charge and $a$ being 1 lead to a Coulombic potential.

Inspired by the energy form, we propose a novel graph convolutional kernel that "pool" neighbor features to update protein $A$'s node features layer by layer:

$$x_{ip}^{(l+1)} = \text{LeakyRelu}\left(\sum_{j \in B} \sum_{k=1}^{K} \left(W_{kp}^{(l+1)} x_i^{(l)}\right)^T A_{ijk} \left(W_{kp}^{(l+1)} x_j^{(l)}\right)\right),$$

where subscript $i$ and $j$ are node indices for proteins $A$ and $B$, respectively, $p$ node-feature index, and $k$ edge-feature index; superscript $(l)$ and $(l+1)$ indicate layer indices. Accordingly, $x_i^{(l)}$ of size $M^{(l)} \times 1$ is the node feature vector for the $i$th node of protein $A$ in layer $l$, $x_{ip}^{(l+1)}$ is the $p$th node-feature for the $i$th node of protein $A$ in layer $(l+1)$ and $W_{kp}^{(l+1)}$ of size $10 \times M^{(l)}$ is the trainable weight matrix for the $k$th edge feature and the $p$th node feature $(p = 1, \ldots, M^{(l+1)})$ in the same layer $(l+1)$. In this way we sum all the interactions between node $i$ in protein $A$ and all neighboring nodes $j$ in protein $B$ through the 11 edge features. Similarly we update node features for each node in protein $B$ by following

$$x_{jp}^{(l+1)} = \text{LeakyReLU}\left(\sum_{i \in B} \sum_{k=1}^{K} \left(W_{kp}^{(l+1)} x_j^{(l)}\right)^T \tilde{A}_{jik} \left(W_{kp}^{(l+1)} x_i^{(l)}\right)\right),$$

where $\tilde{A}_{jik}$ is an element of $\tilde{A}$, a permuted $A$ with the first two dimensions swapped. When calculating intra-molecular energy within a single protein, the second molecule's feature update can be skipped for numerical efficiency.

We only do the convolution on the node features, while keeping the edge features the same across all the EGC layers. The output node-feature matrices $X_A^{(l+1)}$ and $X_B^{(l+1)}$ for both proteins, together with the edge feature tensor $A$, will be used as the input (after batch normalization) for the next layer $(l+1)$. Whereas $M^{(0)} = M = 4$, we choose $M^{(1)} = 2$ and $M^{(2)} = M^{(3)} = 5$ without particular optimization.

*Multi-head attention and fully-connected layers.*

After the 3 EGC layers the output node feature matrices $X_A^{(L)}$ and $X_B^{(L)}$ ($L = 3$ in this study) for both proteins (or two copies of the single matrix in the intra-molecular case) are concatenated and fed into a multi-head attention module [16, 31], whose output subsequently goes through three fully-connected (FC) layers with 128, 64, and 1 output, respectively. A dropout rate of 20% is applied to all but the last FC layer.

*Label, loss function, and model training*

The label here is the binding energy $\Delta G$ although it is not available for encounter complexes. We thus use the same idea as in our previous work [10]. Specifically, using known $k_d$ [32-34], binding affinity of native complexes, or their values predicted from protein sequences [35], we estimate $k_d'$, the binding affinity of an encounter complex, to weaken with the worsening quality of the encounter complex (measured by iRMSD):

$$k_d' = k_d \cdot \exp(\alpha \cdot (iRMSD)^q)$$

where $\alpha$ and $q$ are hyperparameters optimized using the validation set. Specifically, $\alpha$ is searched between 0.5 and 10 with a stepsize of 0.5 and $q$ among 0.25, 0.5, 0.75, 1, 1.5, and 2. The optimized $\alpha$ and $q$ are 1.5 and 0.5, respectively. Therefore, for each sample (corresponding to an encounter complex or decoy), the predicted label is the previously defined $\Delta \hat{G}$, the actual label is $RTln(k'_d) = RTln(k_d) + RT\alpha \cdot (iRMSD)^q$, and the error is simply the difference between the two.

We learn parameters $\boldsymbol{\theta}$ and $\boldsymbol{\theta'}$ of two graph convolutional networks of the same architecture, by minimizing the loss function of mean squared errors (MSE) over samples. Model parameters include $\boldsymbol{W}_{kp}^{(l)}$ in the EGC layers as well as those in the multi-head attention module and FC layers. One set of learned parameter values, $\boldsymbol{\theta}$ are shared among all intra-molecular energy models, and the other set, $\boldsymbol{\theta'}$ are for the inter-molecular energy model.

The model is trained for 50 epochs through the optimizer Adam[36] with a batch size of 16 and the learning rate is tuned to be 0.01 using the validation set.

**Assessment Metrics**

We are aiming at not only relative scoring (ranking) but also absolute scoring (quality estimation) of protein-docking models. Two assessment metrics are therefore introduced.

For ranking models for a given protein-protein pair, we use enrichment factor which is the number of acceptable protein-docking models among the top *P*% ranked by a scoring function divided by that ranked by random. Conceptually, enrichment factor measures the fold improvement of ranking acceptable models high relative to random ranking. Here acceptable models are defined according to the CAPRI criteria with iRMSD within 4 Å.

For quality estimation of all docking models in various quality ranges, we use the root mean square error (RMSE) between the real iRMSD and our predicted iRMSD (as obtained from our predicted labels). Conceptually, RMSE measures the proximity between predicted and actual quality measures.

**Baseline Methods**
We use two baseline scoring functions here. The first one, only applicable to model ranking, is IRAD [12], one of the top-performing scorers in recent CAPRI. The second one is a Random Forest (RF) model from our previous work [10], which both ranks models and estimates model quality. For comparability among models, the RF model is re-trained using the same training and validation sets in this study.

**RESULTS**
**Relative Scoring (Model Ranking)**
We first analyze our model's performances of relative scoring and compare them to IRAD. For both benchmark and CAPRI test sets, we note that the proposed EGCN significantly outperforms both RF and IRAD (Figure 2A and 2B). In particular, for the top-ranked 10% for the benchmark test set, EGCN achieved an average enrichment factor of 3.5, which represents a nearly 40% improvement compared to RF (2.6) and nearly 120% improvement against IRAD (1.6). In all cases, the improvement margins decrease as the top-ranked percentages increase (that is, more models are retained), due to the fact that limited and relatively few acceptable models are available among the total available decoys. Although the training/validation set and the benchmark test set involve unbound docking, the CAPRI test set additionally involves the more challenging cases of homology docking. Nevertheless, for the more challenging CAPRI test set, EGCN's average enrichment factor only slightly dropped from 3.5 to 3.0 in the top-ranked 10% and did so from 3.1 to 2.3 in the top-ranked 20%.

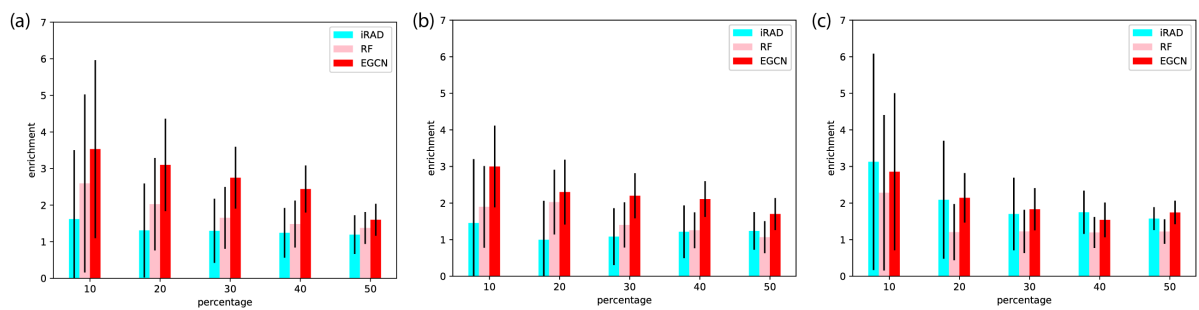

**Figure 2.** Comparing relative scoring (ranking) performances among IRAD, RF, and EGCN. Reported are enrichments ratios of acceptable models among the first *P* percentage, top-ranked decoys for (a) benchmark test set, (b) CAPRI test set, and (c) Score_set, a CAPRI benchmark for scoring.

For Score_set, the CAPRI historical benchmark set, EGCN outperforms RF as shown in Figure 2C. Its average performance is comparable to IRAD although it has considerably lower variance across targets (Figure 2C). It is noteworthy that, although both the benchmark and CAPRI test decoys are generated by the same protocol (ClusPro + cNMA) as the training/validation decoys, Score_set represents completely different and heterogeneous decoy-generating protocols from the community. Impressively, for the top-ranked 10%, EGCN's average enrichment factor is almost the same (nearly 3.0) for Score_set as it is for the CAPRI test set. For the top-ranked 20%, the factor slightly decreased to 2.1 for Score_set compared to 2.3 for the CAPRI test set.

**Absolute Scoring (Quality Estimation)**

We next analyze EGCN's absolute scoring performances and compare them to our previous RF model. EGCN significantly outperforms RF in quality estimation across all test sets (Figure 3). For the benchmark test set, EGCN estimates all docking models' interface RMSD values with an error of 1.32 Å, 1.43 Å, and 1.51 Å when the models' iRMSD values are within 4 Å (acceptable), between 4 Å and 7 Å, and between 7 Å and 10 Å, respectively. These values represent 14%-22% improvement against RF's iRMSD prediction errors. Both EGCN and RF's prediction errors remain relatively flat when models are acceptable or close with iRMSD values within 10 Å whereas they rise out the range when precise quality estimation is no longer desired for those far incorrect models. This performance trend is partially by chance, reflecting the quality distribution in the training set. But it can be guaranteed by design through re-weighting training decoys of different quality ranges, as we did before [10].

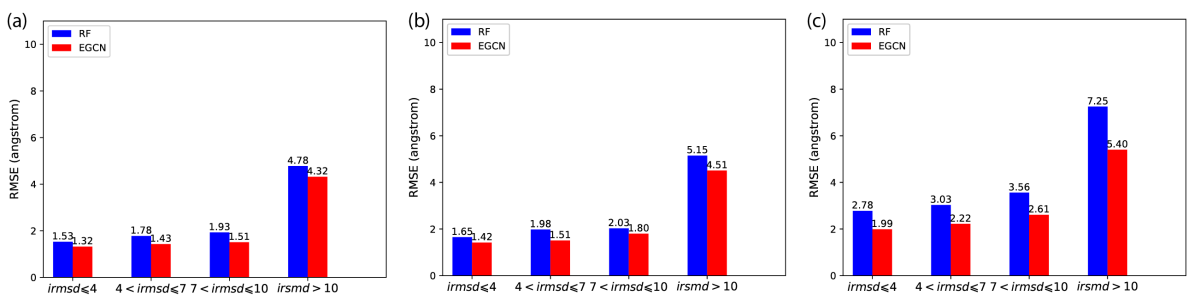

**Figure 3.** Comparing absolute scoring (quality estimation) performances among RF and EGCN. Reported are the RMSE of iRMSD predictions for (a) benchmark test set, (b) CAPRI test set, and (c) Score_set, a CAPRI benchmark for scoring.

Very similar performances and trends are observed for the CAPRI test set and the CAPRI Score_set. EGCN's quality estimation performances only deteriorates slightly in the more challenging CAPRI test set involving homology docking: an error of 1.42 Å, 1.51 Å, and 1.80 Å when the models' iRMSD values are within 4 Å, between 4 Å and 7 Å, and between 7 Å and 10 Å, respectively. For Score_set involving diverse and distinct decoy generation protocols, EGCN's quality-estimation performances further deteriorates slightly to an error of 1.99 Å, 2.22 Å, and 2.61 Å when the models' iRMSD values are within 4 Å, between 4 Å and 7 Å, and between 7 Å and 10 Å, respectively, which has shown even more improvements (27%-28%) relative to RF. It is noteworthy that these two test sets are more challenging than the benchmark test set for additional reasons. Unlike the case of relative scoring, binding affinities of native complexes are needed for absolute scoring. But their values have to be predicted for all but one target pairs in these two test sets (13 out of 14 or 12 out of 14), compared to just 34 out of 107 for the benchmark test set.

Taken together, these results suggest that, learning energy models directly from structures as in EGCN represents a much more accurate and robust data-driven approach than doing so from structure-derived energy terms as in RF. The EGCN model performances are less sensitive to target difficulty than they are to the training versus test data distributions (reflected in the ways training/test decoys are generated).

## DISCUSSION
The accuracy of the EGCN model could be further improved along a few directions. The first and the foremost important direction is to provide better-quality training data whose distributions align better to those of the test data. Specifically in our case, this demands more diverse protocols to generate training decoys (as opposed to the only ClusPro + cNMA protocol used in the study). Moreover, decoys in different quality ranges could be given different priorities as well. The second direction is to train deeper models (going beyond $L = 3$ in this study) and to include more edge features (for instance, including distances besides their reciprocals so that internal energies polynomial in distances can be captured). The third direction is to split the two objectives of relative and absolute ranking. We are training just one model for both model ranking and quality estimation whereas the loss function is perfectly aligned to quality estimation alone. Instead, a separate model with its own loss function can be trained just for model ranking.

The computational cost of training the EGCN model can also be reduced. Currently, we use an edge feature tensor $A \in R^{N \times N \times K}$ and pool all other nodes' features to update each node's feature vector. However, due to the sparsity of protein contact graphs, the edge feature tensor can enjoy a sparse representation of $A \in R^{N \times N' \times K}$ where $N'$ is the maximum number of node neighbors and $N' \ll N$. Here two residues are neighbors if any of the 11 atom-pair distances is below the threshold (12 Å in this study). Accordingly, when updating the feature vector for node $i$, i.e. $x_i^{(l+1)}$, we just need to sum over all its neighboring nodes $j$ for an equivalent expression.

## CONCLUSION

In this paper, we propose a novel, energy-based graph convolutional network (EGCN) for scoring protein-docking models in both relative and absolute senses: ranking docking models and estimating their quality measures (iRMSD). We represent a protein or an encounter complex as an intra-molecular or inter-molecular residue contact graph with atom-resolution node features and edge features. Inspired by physics, we design a novel graph convolutional kernel that maps the inputs (current node features and fixed edge features) to energy-like outputs (next node features). Using such energy-based graph convolution layers and state-of-the-art attention mechanisms, we train two graph convolutional networks of identical architecture and distinct parameters to predict intra- and inter-molecular free-energy, respectively. The two networks are trained to together predict an encounter complex's energy value whereas the true value is approximated by the corresponding native complex's energy discounted according to the quality of the encounter complex.

The first GCN development for protein docking, our EGCN model is tested against three data sets with increasing difficulty levels: unbound docking with a single decoy-generating mechanism, unbound and homology docking with the single mechanism, as well as unbound and homology docking with diverse decoy-generating mechanisms from the community. For ranking protein-docking models (decoys), EGCN is found to perform better or equally well compared to a state-of-the-art method (IRAD) that has found great success in CAPRI as well as our previous RF model[10]. For quality estimation which has seen few method developments in the field, EGCN is again found to outperform our previous RF model. In both cases, EGCN is relatively insensitive to target difficulty and is of limited sensitivity to training decoys (specifically, the way they are generated). Compared to our previous RF model that learns indirectly from structure-derived energy terms, the EGCN model learns directly from structures and thus shows improved accuracy and robustness.

## ACKNOWLEDGEMENTS

This work was supported by the National Institutes of Health (R35GM124952). Portions of this research were conducted with high performance research computing resources provided by Texas A&M University.

# TABLE AND FIGURE LEGENDS

**Table I.** The atom pairs whose distance are converted to edge features. Atom names follow the convention in CHARMM, except that "SC" corresponds to pseudo-atoms for side chains. Note that the last two features are set at 0 for pairs involving prolines (without HN).

**Figure 1.** The architecture of the proposed graph convolutional network (GCN) models for intra- or inter-molecular energies. In our work, there are five types of such models together for predicting encounter-complex binding energy, including 4 intra-molecular models with shared parameters for the unbound or encountered receptor or ligand as well as 1 inter-molecular model for the encounter complex. In each type of model, the inputs (to the left of the arrow) include a pair of node-feature matrices ($X_A$ and $X_B$) for individual protein(s) and an edge-feature tensor ($A$) for intra- or inter-molecular contacts. And the inputs are fed through 3 layers of our energy-based graph convolution layers that learn from training data to aggregate and transform atomic interactions, followed by multi-head attention module and fully-connected layers for the output of intra- or inter-molecular energy.

**Figure 2.** Comparing relative scoring (ranking) performances among IRAD, RF, and EGCN. Reported are enrichments ratios of acceptable models among the first $P$ percentage, top-ranked decoys for (a) benchmark test set, (b) CAPRI test set, and (c) Score_set, a CAPRI benchmark for scoring.

**Figure 3.** Comparing absolute scoring (quality estimation) performances among RF and EGCN. Reported are the RMSE of iRMSD predictions for (a) benchmark test set, (b) CAPRI test set, and (c) Score_set, a CAPRI benchmark for scoring.

# Energy-based Graph Convolutional Networks for Scoring Protein Docking Models
(**Supporting Information**)

| Training Set | 1N8O 7CEI 1DFJ 1AVX<br>1AY7 1IQD 1CGI 1EZU<br>1JPS 1R0R 2FD6 2I25<br>2B42 1EAW 2JEL 1PPE<br>1BJ1 1KXQ 1BVN 1EWY<br>1KAC 1OPH 1E6J 2HLE<br>1WEJ 1A2K 1RLB 1GLA<br>1E6E 1J2J 1BUH 1XQS<br>1M10 1IJK 2HRK 1GP2<br>1GRN 1IBR 1BKD 1Y64 |
|---|---|
| Validation Set | 1ATN 1ACB 1AHW 1AK4<br>1AKJ 1AZS 1B6C 1BVK<br>1XU1 4CPA |
| Benchmark Test Set | 1DQJ 1E4K 1E96 1EER<br>1EFN 1F34 1F51* 1F6M<br>1FAK 1FC2 1FCC* 1FFW<br>1FLE 1FQ1* 1FQJ 1FSK<br>1GCQ 1GHQ 1GL1 1GPW<br>1H9D 1HCF 1HE1* 1HIA<br>1I2M 1I4D 1I9R* 1IB1<br>1IRA* 1JIW 1JK9* 1JMO<br>1JTG 1JWH 1JZD* 1K4C*<br>1K5D 1K74* 1KLU 1KTZ<br>1KXP 1LFD 1MLC 1MQ8<br>1NCA 1NSN 1NW9 1OC0<br>1OFU* 1OYV* 1PVH 1PXV<br>1QA9 1QFW* 1R6Q 1R8S<br>1RV6 1S1Q 1SBB 1SYX*<br>1TMQ* 1UDI* 1US7 1VFB<br>1WQ1 1XD3 1Z0K 1Z5Y*<br>1ZHH* 1ZHI 1ZLI 2A5T*<br>2A9K 2ABZ 2AYO* 2B4J<br>2BTF 2CFH* 2G77* 2H7V*<br>2HQS 2I9B 2IDO* 2J0T<br>2J7P* 2MTA 2NZ8* 2O3B<br>2O8V* 2OOB 2OT3 2OUL<br>2OZA 2PCC 1CLV* 1D6R*<br>2SIC 2SNI 2UUY 2VIS<br>2Z0E* 3CPH 3D5S* 3SGQ* |
|  | 1YVB 9QFW* BOYV* |
| CAPRI Test Set | 2REX* 2WPT 3BX1* 3FM8* 3Q87* 4G9S* 4JW2* 4JW3* 4OJK* 4QKO* 4UEM* 4UF5* 4UHP* 4XL5* |
| Score_set, CAPRI Benchmark Test Set | 2VDU* 2R20* 3BX1* 2W5F* 2W83* 3FM8* 3FM8* 3E8L* 2WPT 3Q87* 3U43* 4JW2* 4JW3* |

**Table S1**. Protein complexes (represented by their 4-letter PDB IDs) used in training, validation, and 3 test sets. The PDB IDs with stars correspond to cases where native binding affinities were predicted from protein sequences.

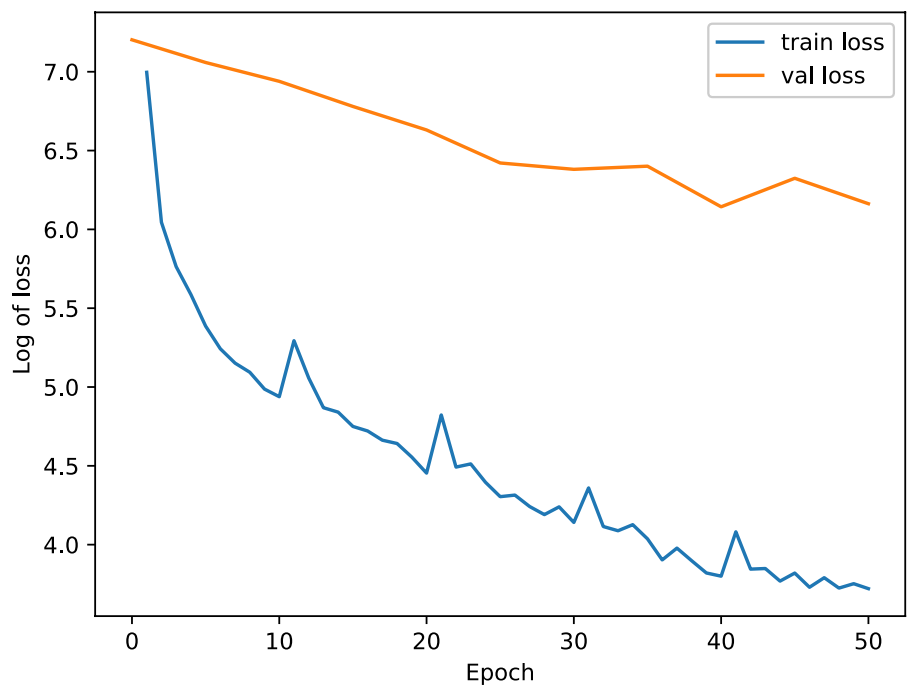

**Figure S1**. Training and validation loss over epochs during the training of the EGCN model.